\documentclass[pra,12pt,tightenlines,superscriptaddress,eqsecnum,secnumarabic]{revtex4}

\usepackage{bm,amssymb,amsmath,amsthm,mathrsfs,multirow,longtable,url,colonequals,verbatim,afterpage,ifmtarg}

\usepackage[continuous]{pagenote}
\makepagenote

\usepackage{hyperref}
\makeatletter
\@ifundefined{pdfbookmark}{\def\pdfbookmark[#1]#2#3{}}{}
\makeatother

\theoremstyle{plain}
\newtheorem{thm}{Theorem}[section]
\newtheorem{con}[thm]{Conjecture}

\DeclareSymbolFont{AMSb}{U}{msb}{m}{n}
\DeclareMathSymbol{\N}{\mathbin}{AMSb}{"4E}
\DeclareMathSymbol{\Z}{\mathbin}{AMSb}{"5A}
\DeclareMathSymbol{\R}{\mathbin}{AMSb}{"52}
\DeclareMathSymbol{\Q}{\mathbin}{AMSb}{"51}
\DeclareMathSymbol{\I}{\mathbin}{AMSb}{"49}
\DeclareMathSymbol{\C}{\mathbin}{AMSb}{"43}
\DeclareMathSymbol{\F}{\mathbin}{AMSb}{"46}
\DeclareMathSymbol{\E}{\mathbin}{AMSb}{"45}
\DeclareMathSymbol{\K}{\mathbin}{AMSb}{"4B}
\DeclareMathSymbol{\LL}{\mathbin}{AMSb}{"4C}

\def\coloneq{\colonequals}
\def\eqcolon{\equalscolon}

\def\ket#1{{|#1\rangle}}
\def\bra#1{{\langle#1|}}
\def\ketbra#1{{|#1\rangle\langle#1|}}
\def\braket#1#2{{\langle#1|#2\rangle}}

\def\spf#1#2{{\langle#1,#2\rangle}}
\def\md#1{{\:(\operatorname{mod}\,#1)}}

\def\tr{{\operatorname{tr}}}

\def\d{\mathrm{d}}
\def\U{{\operatorname{U}}}

\def\PC{{\operatorname{PC}}}
\def\PEC{{\operatorname{PEC}}}
\def\ECl{{\operatorname{EC}}}
\def\Cl{{\operatorname{C}}}
\def\cl{{\operatorname{c}}}
\def\ecl{{\operatorname{ec}}}
\def\GL{{\operatorname{GL}}}
\def\SLtwo{{\operatorname{SL}_2}}
\def\ESLtwo{{\operatorname{ESL}_2}}
\def\H{{\operatorname{H}}}
\def\I{{\operatorname{I}}}

\def\PCd{{\C P^{d-1}}}
\def\db{{\bar{d}}}
\def\Fz{{F_z}}
\def\Sz{{Z}}
\def\Fsa{{F_a}}
\def\Fsb{{F_b}}
\def\Fsc{{F_c}}
\def\orb{{\operatorname{orb}}}
\def\stab{{\operatorname{S}}}

\def\mr#1#2{\multirow{#1}{*}{#2}}
\def\mat#1#2#3#4{\bigl(\begin{smallmatrix}{#1}&{#2}\\{#3}&{#4}\end{smallmatrix}\bigr)}

\def\matcv#1#2#3#4#5#6{\bigl(\begin{smallmatrix}{#1}&{#2}\\{#3}&{#4}\end{smallmatrix}\big|\begin{smallmatrix}{#5}\\{#6}\end{smallmatrix}\bigr)}
\def\matbv#1#2#3#4#5#6{\bigl[\begin{smallmatrix}{#1}&{#2}\\{#3}&{#4}\end{smallmatrix}\big|\begin{smallmatrix}{#5}\\{#6}\end{smallmatrix}\bigr]}
\def\ce#1#2{{({#1}\mspace{1mu}|\mspace{1mu}{#2})}}
\def\cle#1#2{{[{#1}\mspace{1mu}|\mspace{1mu}{#2}]}}

\def\printfid#1{\pagebreak[0]\begin{samepage}\noindent\pdfbookmark[2]{#1}{numeric_#1}\textbf{$\bm{#1}$}\penalty10000\begin{scriptsize}\verbatiminput{sicfiducial_#1.txt}\end{scriptsize}\vskip20pt plus 10pt minus 10pt\end{samepage}}

\def\printfidsym#1{\penalty-100\noindent\pdfbookmark[2]{#1}{symbolic_#1}\textbf{$\bm{#1}$}\penalty10000\begin{scriptsize}\verbatiminput{sicsymbolic_#1.txt}\end{scriptsize}\vskip20pt plus 10pt minus 10pt}

\begin{document}

\title{SIC-POVMs: A new computer study}

\author{A. J. Scott}
\email{andrew.scott@griffith.edu.au}
\affiliation{Centre for Quantum Computer Technology and Centre for Quantum Dynamics, Griffith University, Brisbane, Queensland 4111, Australia}
\author{M. Grassl}
\email{markus.grassl@nus.edu.sg}
\affiliation{Centre for Quantum Technologies, National University of Singapore,
3 Science Drive 2, Singapore 117543, Singapore}

\begin{abstract}
We report on a new computer study into the existence of $d^2$
equiangular lines in $d$ complex dimensions. Such maximal complex
projective codes are conjectured to exist in all finite dimensions and
are the underlying mathematical objects defining symmetric
informationally complete measurements in quantum theory. We provide
numerical solutions in all dimensions $d\leq 67$ and, moreover, a
putatively complete list of Weyl-Heisenberg covariant solutions for
$d\leq 50$. A symmetry analysis of this list leads to new algebraic
solutions in dimensions $d=24$, $35$ and $48$, which are given
together with algebraic solutions for $d=4,\ldots,15$ and $19$.
\end{abstract}

\maketitle

\section{Introduction}
\label{sec:intro}

Interest in sets of equiangular lines began at least 60 years ago~\cite{Haantjes48,VanLint66,Lemmens73,Delsarte75,Delsarte77} and continues to this day~\cite{Bannai09}. In this article we report on a new computer study of what has now become one of the most urgent unanswered questions: {\em Do there exist $d^2$ equiangular lines, the maximum possible, in all finite complex dimensions $d$?\/} This question is attracting increasing attention from the quantum physics community~\cite{Renes04,Grassl04,Appleby05,Grassl05,Grassl06,Grassl08a,Grassl08b,Klappenecker05,Colin05,Wootters06,Flammia06,Scott06,Appleby07,Belovs08,Appleby09,Fuchs09,Bengtsson09,Appleby09b,Appleby09c} and, more recently, from the communities of design theory~\cite{Khatirinejad08,Godsil09} and frame theory~\cite{Howard05,Waldron07,Fickus09}. The believed affirmative answer, originally conjectured 10 years ago by Gerhard Zauner~\cite{Zauner99}, has now been confirmed exactly in dimensions $d=2,3$~\cite{Delsarte75}, $4,5$~\cite{Zauner99}, 6~\cite{Grassl04}, 7~\cite{Appleby05}, 8~\cite{Hoggar98,Grassl05} $9,\dots,13,15$~\cite{Grassl05,Grassl06,Grassl08a,Grassl08b} and 19~\cite{Appleby05}, and to high numerical precision in all dimensions $d\leq 45$~\cite{Renes04}. The fundamental question remains unresolved, however, inviting speculation as to whether the true answer is negative, with Zauner's conjecture failing in some large untested dimension, or simply that the affirmative answer is truly difficult to prove, with Zauner's conjecture remaining open for many years to come. In either case, a new computer study is timely. 

\section{SIC-POVMs}
\label{sec:sics}

In quantum theory, a set of $d^2$ equiangular lines in $d$ complex
dimensions is the underlying mathematical object defining a {\em
  symmetric informationally complete positive-operator-valued measure
  (SIC-POVM)\/}~\cite{Renes04}. These measures describe the
measurement-outcome statistics of a particularly attractive choice of
a `standard' informationally complete quantum measurement, both from a
foundational perspective~\cite{Fuchs09} and for the purpose of quantum
state tomography~\cite{Scott06}. In precise terms, a SIC-POVM $P$ is a
{\em positive-operator-valued measure (POVM)\/} (see
e.g.\ ref.~\cite{Busch96}) that maps each of its $d^2$ possible
measurement outcomes, denoted $x_1,\dots,x_{d^2}$ say, to one of $d^2$
subnormalised rank-one projectors on the Hilbert space of
$d$-dimensional pure quantum states $\C^d$,
\begin{align}
P(x_k)\coloneq\frac{1}{d}\,\ketbra{x_k}\:, 
\end{align}
with the defining property that equiangularity is enjoyed under the Hilbert-Schmidt inner product:
\begin{align}\label{eq:sic}
\tr\bigl(P(x_j)P(x_k)\bigr) = \frac{1}{d^2}\,|\braket{x_j}{x_k}|^2 = \frac{d\delta_{j,k}+1}{d^2(d+1)}\:.
\end{align}

Considering the rays in $\C^d$ upon which each $\ketbra{x_k}$ projects, a SIC-POVM is of course equivalent to a set of $d^2$ equiangular lines through the origin of $\C^d$. It is therefore natural to identify the outcome set as a subset of complex projective space, $\mathscr{X}\coloneq\{x_1,\dots,x_{d^2}\}\subset\PCd$. A set of equiangular lines $\mathscr{C}\subset\PCd$, $|\braket{x}{y}|^2=\alpha<1$ for all $x\neq y\in\mathscr{C}$, is then a type of {\em complex projective code\/}, called a {\em $\PCd$ 1-distance set\/} (see e.g.\ refs.~\cite{Rankin55,Conway88,Levenshtein98}). It is known that any such $\mathscr{C}$ obeys the so-called absolute bound on its size: $|\mathscr{C}|\leq d^2$. The maximum requires the same common angle enjoyed by SIC-POVMs. Alternatively, in terms of line packings, for any set $\mathscr{S}\subset\PCd$ of size $|\mathscr{S}|=d^2$, it is known that $\max_{x\neq y\in\mathscr{S}}|\braket{x}{y}|^2\geq 1/(d+1)$ with equality only if $\mathscr{S}$ is equiangular~\cite{Rankin55}.

A dual characterisation of SIC-POVMs comes from design theory (see e.g.\ refs.~\cite{Levenshtein98,Harpe05,Hoggar82} or sec.\ 2 of ref.~\cite{Roy07} for a concise introduction). A finite set $\mathscr{D}\subset\PCd$ is called a {\em complex projective $t$-design\/} if
\begin{align}\label{eq:tdesign}
\frac{1}{|\mathscr{D}|}\sum_{x\in\mathscr{D}}\ketbra{x}^{\otimes t}&=\mathop{\int}_{\PCd}\d\mu(x)\,\ketbra{x}^{\otimes t}\:,
\end{align}
where $\mu$ is the Haar measure. In these terms, SIC-POVMs are precisely equivalent to {\em tight\/} complex projective 2-designs. These are 2-designs that meet the absolute bound on their size: $|\mathscr{D}|\geq d^2$. All 2-designs with $|\mathscr{D}|=d^2$ are necessarily sets of equiangular lines and these are the only 2-designs with this structure. This characterisation has a straightforward generalisation in terms of {\em weighted $t$-designs\/}~\cite{Levenshtein98} or {\em cubature formulas\/}~\cite{Harpe05}.

A third characterisation of SIC-POVMs is in terms of frame theory (see e.g.\ refs.~\cite{Kovacevic07,Christensen03}). In this context, a set of unit vectors that specifies a SIC-POVM, $\{\ket{x_k}\}_{k=1}^{d^2}\subset\C^d$, is called a maximally equiangular tight frame~\cite{Fickus09}. More importantly, under the projection $P\mapsto P-\tr(P)I/d$ into the real vector space of traceless Hermitian operators (the natural generalisation of the Bloch-sphere representation to higher dimensions~\cite{Scott06,Appleby09}) a SIC-POVM again maps to a tight frame (in this case a simplex), which means the representation 
\begin{align}
\rho &= d(d+1)\sum_{k}\tr\bigl(P(x_k)\rho\bigr)P(x_k)-I 
\end{align}
is afforded by any quantum state $\rho$ (see refs.~\cite{Scott06,Scott08} for descriptions of such {\em tight\/} informationally complete POVMs). This state-inversion formula for $\rho$ in terms of its measurement statistics $\tr\bigl(P(x_k)\rho\bigr)$ immediately proves {\em informational completeness\/}~\cite{Busch91} for SIC-POVMs. Moreover, amongst all {\em minimally\/} informationally complete POVMs (i.e.\ those having $d^2$ outcomes), this representation is unique to SIC-POVMs. These considerations have lead some to argue~\cite{Scott06,Appleby07} that SIC-POVMs should be promoted to the  unique status of {\em standard\/} informationally complete POVMs, being as close as possible to orthonormal bases for the space of quantum states. Indeed, SIC-POVMs would be the best choice in any bid to standardise experimental reporting in quantum state tomography, being the most robust minimally informationally complete POVMs against statistical error~\cite{Scott06}.

\section{Weyl-Heisenberg SIC-POVMs and the Clifford group}
\label{sec:whsics}

The most promising route towards a general construction of SIC-POVMs involves translating a fiducial vector under the {\em Weyl displacement operators\/}~\cite{Weyl50,Schwinger60}:
\begin{align}
D_p &\coloneq \tau^{p_1p_2} V^{p_1} U^{p_2} \:,\qquad V\ket{k}=\ket{k+1\md{d}} \:,\qquad U\ket{k}=\omega^k\ket{k}\:,
\end{align}
where $p=(p_1,p_2)\in\Z^2$, $\tau=e^{\pi i(d+1)/d}$, $\omega=\tau^2=e^{2\pi i/d}$ (meaning $\tau^{d^2}=\tau^{2d}=\omega^d=1$), and we have fixed an orthonormal basis for $\C^d$: $\ket{0},\dots,\ket{d-1}$. Defining the symplectic form 
\begin{align}
\spf{p}{q} &\coloneq p_2q_1-p_1q_2\:, 
\end{align}
these operators obey the relations
\begin{align}
D_p D_q &= \tau^\spf{p}{q}D_{p+q} \label{eq:disprel1}\\ 
{D_p}^\dag &= D_{-p} \\
D_{p+dq} &= \begin{cases} D_p\:, & \text{if $d$ is odd;}\\ (-1)^\spf{p}{q}D_p\:, & \text{if $d$ is even,}  \end{cases}
\end{align}
and together generate a variant of the {\em Heisenberg group\/}:
\begin{align}
\H(d)\coloneq\{ e^{i\xi} D_p: p\in\Z^2,\xi\in\R \}\:.
\end{align}
Modulo its center, $\I(d)\coloneq\{e^{i\xi}I:\xi\in\R\}$, the Heisenberg group is simply a direct product of cyclic groups, $\H(d)/\I(d)\cong{\Z_d}^2$, where $\Z_d=\Z/d\Z=\{0,\dots,d-1\}$.

It was conjectured in ref.~\cite{Renes04} that, in every finite dimension, a SIC-POVM can be constructed as the orbit of a suitable {\em fiducial vector\/} $\ket{\phi}\in\C^d$ under the action of the displacement operators: 
\begin{align}
P(x_{p_1+p_2d+1})=\frac{1}{d}\,D_p\ketbra{\phi}{D_p}^\dag\:.
\end{align}
The condition for equiangularity (\ref{eq:sic}) then becomes
\begin{align}\label{eq:whsic}
|\bra{\phi}D_p\ket{\phi}|^2 = \frac{d\delta_{p,0}+1}{d+1}\:.
\end{align}
To bolster this conjecture, such {\em Weyl-Heisenberg covariant SIC-POVMs\/} were found with high numerical precision in all dimensions $d\leq 45$. Unbeknownst to the authors of ref.~\cite{Renes04}, however, a stronger conjecture had already been put forward by Gerhard Zauner in his doctoral dissertation~\cite{Zauner99}. Zauner claimed that, in every finite dimension, a fiducial vector for a Weyl-Heisenberg covariant SIC-POVM can be found in an eigenspace of the matrix\pagenote{In his doctoral dissertation \cite[p.~65]{Zauner99}, following the derivation of SIC-POVMs for dimensions $2,\ldots,7$, Gerhard Zauner makes the following observation:
\begin{quote}
Die Beispiele f\"ur $b=2,3,4,5,6,7$ legen die folgende Vermutung
nahe. F\"ur alle $b\ge 2$ gibt es im
($\left[\frac{b}{3}\right]+1$)-dimensionalen Eigenraum zum Eigenwert
$1$ der $b\times b$ Matrix $\mathbf{Z}$ Vektoren, welche mit dem
Ansatz (3.14) maximale Quantendesigns mit Grad $1$ erzeugen.  F\"ur
$b=3m+2$ gibt es im gleichdimensionalen Eigenraum zum Eigenwert
$\alpha$ ebensolche Vektoren.
\end{quote}
In the present context, this quotation can be translated as follows ($\mathbf{Z}$ is the transpose of $\Sz$, in fact, but we will follow Appleby~\cite{Appleby05}):
\begin{quote}
The examples for dimension $d=2,3,4,5,6,7$ give rise to the following
conjecture. For all dimensions $d\ge 2$ the eigenspace of dimension
$\lfloor\frac{d}{3}\rfloor+1$ with eigenvalue $1$ of the $d\times d$
matrix $Z$ (see eq. (\ref{eq:Zmatrix})) contains fiducial vectors of a
Weyl-Heisenberg covariant SIC-POVM. For $d=3m+2$ the eigenspace of
the same dimension with eigenvalue $e^{2\pi i/3}$ contains
fiducial vectors as well.
\end{quote}} 
\begin{align}\label{eq:Zmatrix}
\bra{j}\Sz\ket{k}&\coloneq \frac{e^{i\xi}}{\sqrt{d}}\,\tau^{2jk+j^2}\:.
\end{align} 
\begin{con}[Zauner~\cite{Zauner99}]
In all finite dimensions there exists a fiducial vector for a Weyl-Heisenberg covariant SIC-POVM that is an eigenvector of $\Sz$.
\end{con}
Setting $\xi=\pi(d-1)/12$, it can be shown~\cite{Zauner99} that $\Sz$ has order 3: $\Sz^3=I$. The eigenspace with eigenvalue $e^{2\pi ik/3}$ will be labeled $\mathcal{Z}_k$ ($k=0,1,2$). Then
\begin{align}\label{eq:Zeignenspaces}
\dim \mathcal{Z}_k &= \lfloor (d+3-2k)/3\rfloor \:.
\end{align}

Under the action of conjugation, $\Sz$ defines an automorphism of the
Heisenberg group, $\Sz^{-1}\H(d)\Sz=\H(d)$, and therefore belongs to
the normaliser of $\H(d)$ in $\U(d)$,
\begin{align}
\Cl(d) &\coloneq\{U\in\U(d):U^{-1}\H(d)U=\H(d)\} \:,
\end{align}
which is called the {\em Clifford group\/} in quantum information theory, but more widely recognised as a variant of the {\em Jacobi group\/}~\cite{Berndt98}. The significance of $\Cl(d)$ to SIC-POVMs follows from eq.~(\ref{eq:whsic}): if $\ket{\phi}$ is a fiducial vector for a Weyl-Heisenberg covariant SIC-POVM, then so is $U\ket{\phi}$ for any $U\in\Cl(d)$.   

An explicit description of the Clifford group can be easily deduced in odd dimensions, which we now summarise. In general dimensions, for each symplectic matrix $F\in\SLtwo(\Z_d)$ and $p\in{\Z_d}^2$, let 
\begin{align}\label{eq:Fpnotation}
\ce{F}{p}=\left(\begin{array}{cc|c} F_{11} & F_{12} & p_1 \\ F_{21} & F_{22} & p_2 \end{array}\right)\coloneq\begin{pmatrix} F_{11} & F_{12} & p_1 \\ F_{21} & F_{22} & p_2 \\ 0 & 0 & 1 \end{pmatrix} ,
\end{align}
and define the matrix group
\begin{align}\label{eq:cstruct}
\cl(d) &\coloneq  \bigl\{\ce{F}{p}:F\in\SLtwo(\Z_d),p\in{\Z_d}^2\bigr\}\cong \SLtwo(\Z_d)\ltimes{\Z_d}^2\:,
\end{align}
which follows the multiplication rule
\begin{align}
\ce{F}{p}\ce{G}{q} &= \ce{FG}{p+Fq} \:.
\end{align}

Now assume $d$ is odd. Then for each $\ce{F}{p}\in\cl(d)$ there is a unique unitary $C_{\ce{F}{p}}\in\Cl(d)$, up to the multiplication of a phase $e^{i\xi}$, for which
\begin{align}\label{eq:metaconjugation}
C_{\ce{F}{p}}D_q{C_{\ce{F}{p}}}^\dag &= \omega^\spf{p}{Fq}D_{Fq}
\end{align}
for all $q\in\Z^2$. All Clifford operators take this action and, in fact, $C_g$ is a faithful projective unitary representation of $\cl(d)$ (i.e.\ $C_gC_h=e^{i\xi(g,h)}C_{gh}$ for some $\xi:\cl(d)\times\cl(d)\rightarrow\R$) 
that defines the group isomorphism
\begin{align}
\cl(d) &\cong \Cl(d)/\I(d)\:.
\end{align}
The Clifford group describes all automorphisms of $\H(d)$ that leave its center pointwise fixed: the inner automorphisms are just displacement operators, $D_p=C_{\ce{I}{p}}$, while the outer automorphisms are specified by the operators $C_{\ce{F}{0}}$, which are more widely recognised as {\em metaplectic operators\/}~\cite{Weil63,Feichtinger08}. Since eq.~(\ref{eq:metaconjugation}) defines each Clifford operator uniquely, up to a phase $e^{i\xi}$, it can be used to derive formulas, e.g., by multiplying on the right by ${D_{q}}^\dag$ and summing over $q$ to obtain  
\begin{align}\label{eq:oddCdefn}
C_{\ce{F}{p}} &= \frac{e^{i\xi}}{d\sqrt{\eta(F)}}\,D_{p}\sum_r D_{Fr}{D_{r}}^\dag\:,
\end{align}
where $\eta(F)\coloneq |\{q\in{\Z_d}^2|Fq=q\}|=|\tr\,C_{\ce{F}{0}}|^2$. To check eq.~(\ref{eq:metaconjugation}), simply replace $r$ by $r+q$ in the sum, and use $\spf{Fr}{Fq}=(\det F)\spf{r}{q}$ and eq.~(\ref{eq:disprel1}) repeatedly to obtain:
\begin{align}
C_{\ce{F}{p}} &= \omega^\spf{p}{Fq}D_{Fq}\,\frac{e^{i\xi}D_{p}}{d\sqrt{\eta(F)}}\sum_r\tau^{(\det F-1)\spf{r}{q}} D_{Fr}{D_{r}}^\dag\,{D_{q}}^\dag \\
&= \omega^\spf{p}{Fq}D_{Fq}C_{\ce{F}{p}}{D_{q}}^\dag\:,
\end{align}
since $\det F=1 \md{d}$ and $\tau^d=1$ for odd $d$. In even dimensions we would need $\det F=1 \md{2d}$, however, or we would need to generalise eq.~(\ref{eq:metaconjugation}) so that the factor $\tau^{(\det F-1)q_1q_2}$ appears on the right. Both possibilities indicate how the above description of $\Cl(d)$ might be amended to work in general. The former choice was taken by Appleby~\cite{Appleby05} and the latter by Feichtinger {\it et al.\/}~\cite{Feichtinger08}. We will follow Appleby's approach, which we now summarise. 

In even dimensions, the metaplectic operators can introduce sign changes to eq.~(\ref{eq:metaconjugation}) unless we instead take $F\in\SLtwo(\Z_{2d})$, as done by Appleby~\cite{Appleby05}. We also take $p\in{\Z_{2d}}^2$ for simplicity. In general dimensions, let
\begin{align}
\db &\coloneq \begin{cases} d\:, & \text{if $d$ is odd;}\\ 2d\:, & \text{if $d$ is even,}  \end{cases}
\end{align}
and identify
\begin{align}
F &= \begin{pmatrix} \alpha & \beta \\ \gamma & \delta \end{pmatrix} \in \SLtwo(\Z_{\db}) \:.
\end{align}
Now, and hereafter, for each $\ce{F}{p}\in\cl(\db)$ define
\begin{align}\label{eq:genCdefn}
C_{\ce{F}{p}} &\coloneq D_p V_F\:,
\end{align}
with
\begin{align}\label{eq:Vdefn}
\bra{j}V_F\ket{k} \coloneq \frac{1}{\sqrt{d}}\,\tau^{\beta^{-1}\left(\alpha k^2-2jk+\delta j^2\right)}\,
\end{align}
if there exists an element $\beta^{-1}\in\Z_{\db}$ with $\beta^{-1}\beta=1\md{\db}$; otherwise, we find an integer $x\in\Z_{\db}$ with the property that $(\delta+x\beta)^{-1}\in\Z_{\db}$ (whose existence is guaranteed~\cite[Lemma~3]{Appleby05}) and take  $V_F=V_{F_1}V_{F_2}$, using eq.~(\ref{eq:Vdefn}) now for $V_{F_1}$ and $V_{F_2}$, where
\begin{align}
F_1 &= \begin{pmatrix} 0 & -1 \\ 1 & x \end{pmatrix} \quad\text{and}\quad F_2 = \begin{pmatrix} \gamma+x\alpha & \delta+x\beta \\ -\alpha & -\beta \end{pmatrix}
\end{align}
fulfill the decomposition $F=F_1F_2$. In odd dimensions, eq.~(\ref{eq:genCdefn}) is equivalent to eq.~(\ref{eq:oddCdefn}). In all dimensions, now choosing $\xi=0$ in eq.~(\ref{eq:Zmatrix}), Zauner's matrix is $\Sz=C_{\ce{\Fz}{0}}$, where
\begin{align}\label{eq:Fz}
\Fz &\coloneq \begin{pmatrix} 0 & d-1 \\ d+1 & d-1 \end{pmatrix}.
\end{align}

With these definitions, Appleby~\cite[Theorem~1]{Appleby05} showed that the map $g\mapsto C_g$ defines a group isomorphism 
\begin{align}\label{eq:Cstruct}
\cl(\db)/\ker(C) \cong\Cl(d)/\I(d) \eqcolon \PC(d)\:, 
\end{align}
obeying eq.~(\ref{eq:metaconjugation}), where
\begin{align}
\ker(C) &= \begin{cases} \bigl\{ \matcv{1}{0}{0}{1}{0}{0} \bigr\}\:, & \text{if $d$ is odd;}\\ \bigl\langle\matcv{1+d}{0}{0}{1+d}{0}{0},\matcv{1}{d}{0}{1}{d/2}{0},\matcv{1}{0}{d}{1}{0}{d/2}\bigr\rangle\:, & \text{if $d$ is even.}  \end{cases}
\end{align}
Combining eqs.~(\ref{eq:cstruct}) and (\ref{eq:Cstruct}) we can deduce the size of the Clifford group. It is known that
\begin{align}
|\SLtwo(\Z_d)|=d^3\prod_{p\mid d}\bigl(1-p^{-2}\bigr)\:,
\end{align}
where the product is over all primes $p$ dividing $d$, which means $|\cl(\db)|=32|\cl(d)|$ for even $d$. But since $|{\ker(C)}|=32$ in even dimensions, we can conclude that
\begin{align}
|\PC(d)|=|\cl(d)|=|\SLtwo(\Z_{d})||{\Z_{d}}^2|=d^5\prod_{p\mid d}\bigl(1-p^{-2}\bigr)
\end{align}
in all dimensions.

Finally, note that there is a complex-conjugation symmetry apparent in eq.~(\ref{eq:whsic}). Let $\hat{J}=\hat{J}^\dag$ be the anti-unitary operator with action $\hat{J}\sum_k c_k\ket{k}=\sum_k {c_k}^*\ket{k}$ for a vector rewritten in our standard basis. Then, 
\begin{align}
\hat{J} D_p \hat{J}^\dag &= D_{Jp}
\end{align}
for all $p\in\Z^2$, where
\begin{align}
J &\coloneq \begin{pmatrix} 1 & 0 \\ 0 & -1 \end{pmatrix}.
\end{align}
It thus follows from eq.~(\ref{eq:whsic}) that $\hat{J}\ket{\phi}$ is a fiducial vector for a Weyl-Heisenberg covariant SIC-POVM whenever $\ket{\phi}$ is. To analyze this additional symmetry, define the matrix group
\begin{align}\label{eq:ecstruct}
\ecl(d) &\coloneq  \bigl\{\ce{F}{p}:F\in\ESLtwo(\Z_d),p\in{\Z_d}^2\bigr\}\cong \ESLtwo(\Z_d)\ltimes{\Z_d}^2\:,
\end{align}
where $\ESLtwo(\Z_d)\coloneq\bigl\{F\in{\operatorname{Mat}_{2,2}}(\Z_d):\det F=\pm 1\md{d}\bigr\}=\SLtwo(\Z_d)\cup J\,\SLtwo(\Z_d)$ is the union of all symplectic and anti-symplectic matrices. The {\em extended Clifford group\/} is then defined as the group of all unitary and anti-unitary operators that normalise $\H(d)$, i.e., the disjoint union 
\begin{align}\label{eq:ECdefn}
\ECl(d) &\coloneq \Cl(d)\cup \hat{J}\,\Cl(d) \:.
\end{align}
Appleby~\cite[Theorem~2]{Appleby05} showed that 
\begin{align}\label{eq:ECstruct}
\ecl(\db)/\ker(E) \cong\ECl(d)/\I(d) \eqcolon \PEC(d)\:, 
\end{align} 
through the map $E:\ecl(\db)\rightarrow\ECl(d)$, where
\begin{align}\label{eq:genEdefn}
E_{\ce{F}{p}} &\coloneq  \begin{cases} C_{\ce{F}{p}}\:, & \text{if $\ce{F}{p}\in\cl(\db)$;}\\ \hat{J}C_{\ce{JF}{Jp}}\:, & \text{otherwise.}  \end{cases}
\end{align}
From eq.~(\ref{eq:ECdefn}), we know that $\ker(E)=\ker(C)$ and thus have $|\PEC(d)|=2|\PC(d)|$. 

Lastly, in eqs.~(\ref{eq:Cstruct}) and (\ref{eq:ECstruct}) we have
defined projective versions of the Clifford and extended Clifford
groups, respectively. The notation $[U]\coloneq
\{e^{i\xi}U\}_{\xi\in\R}$ will be used in the following section to
denote members of these groups, which are equivalence classes of
unitary and anti-unitary matrices that differ only by a factor of unit modulus.  We will also use the shorthand notation
\begin{align}
\cle{F}{p} &= \left[\begin{array}{cc|c} F_{11} & F_{12} & p_1 \\ F_{21} & F_{22} & p_2 \end{array}\right] \coloneq [E_{\ce{F}{p}}]\in\PEC(d) \:.
\end{align}

\section{Numerical computer solutions}
\label{sec:numer}

The task of finding SIC-POVMs is facilitated by the following recharacterisation of $t$-designs (see e.g.\ ref.~\cite{Roy07}): for any finite $\mathscr{S}\subset\PCd$, and any positive integer $t$,
\begin{align}\label{eq:welchbound}
\frac{1}{|\mathscr{S}|^2}\sum_{x,y\in\mathscr{S}}\,|\braket{x}{y}|^{2t} &\geq \tbinom{d+t-1}{t}^{-1}\:,
\end{align}
with equality if and only if $\mathscr{S}$ is a $t$-design. This allows us to check whether a proposed set $\mathscr{S}\subset\PCd$ forms a $t$-design by considering only the angles between members. It also shows that $t$-designs can be found numerically by parameterising $\mathscr{S}$ and minimising the LHS of eq.~(\ref{eq:welchbound}). The lower bound was derived by Welch~\cite{Welch74}. Setting $\ket{x_{p_1+p_2d+1}}=D_p\ket{\phi}$, $|\mathscr{S}|=d^2$ and $t=2$, we can translate eq.~(\ref{eq:welchbound}) to our case: {\it for any\/ $\ket{\phi}\in\C^d$,
\begin{align}\label{eq:sicbound}
\frac{1}{d}\sum_p|\bra{\phi}D_p\ket{\phi}|^4=\sum_{j,k}\Big|\sum_l\braket{\phi}{j+l}\braket{l}{\phi}\braket{\phi}{k+l}\braket{j+k+l}{\phi}\Big|^2 &\geq \frac{2}{d+1}\:,
\end{align}
with equality if and only if\/ $\ket{\phi}$ is a fiducial vector for a Weyl-Heisenberg covariant SIC-POVM.\/} The second form of this sum was used to obtain the numerical results reported next.\pagenote{Solutions for fiducial vectors were found by minimising the LHS of eq.~(\ref{eq:sicbound}), the cost function, until the bound on the RHS was met. This was performed using the MATLAB\textsuperscript{\textregistered} Optimization Toolbox\texttrademark~\cite{MATLAB} with the cost function and its derivatives implemented in C and linked in. The solutions were then further refined to 38 digits with the multiprecision capabilities of PARI/GP~\cite{PARIGP}. A small number of AMD Opteron\texttrademark~252 dual-processor machines were used, though for a significant amount of time.}

Our extensive computer searches discovered fiducial vectors in all dimensions $d\leq 67$, improving considerably on the solutions reported previously~\cite{Renes04}. These are listed in appendix~\ref{app:numerics} and may be assumed exact to the 38 digits quoted. Note that each solution $\phi$ will generate an entire orbit of related fiducial vectors under the action of the extended Clifford group:
\begin{align}
\orb(\phi) &\coloneq \{U\ketbra{\phi}U^{\dag}\}_{U\in\ECl(d)}\:.
\end{align}
Those listed in the appendix generate unique orbits. Moreover, we are confident that this list is complete for $d\leq 50$, where the computer search was exhaustive.\pagenote{In each dimension $d$, the current list of known $\PEC(d)$ orbits of fiducial vectors is considered complete when, upon initialising each trial to a random vector under the Haar measure, the search consecutively encounters $30(n+1)$ solutions that generate one of the $n$ known orbits. Assuming each orbit is found with equal probability, the probability of missing an orbit is then no more than $e^{-30}$. Under this criteria, the list is complete for $d\leq 47$. In dimensions $d=48,\dots,50$, we have encountered enough of the known orbits to be confident that our list is complete here also, but the computations are ongoing.}

In table~\ref{tbl:symmetries} we list the total number of unique Weyl-Heisenberg SIC-POVMs in each dimension in terms of orbits of fiducial vectors. Note that the length of each orbit will be $|\PEC(d)|$, unless there is a symmetry present in $\phi$, described by its stabiliser 
\begin{align}
\stab(\phi) &\coloneq \{[U]\in\PEC(d):|\bra{\phi}U\ket{\phi}|=1\}\:,
\end{align}
where $[U]\coloneq \{e^{i\xi}U\}_{\xi\in\R}$. We then have $|\orb(\phi)|=|\PEC(d)|/|\stab(\phi)|$, giving
\begin{align}
\text{\# SIC-POVMs} &= \frac{|\PEC(d)|}{d^2} \sum_{\orb(\phi)}\frac{1}{|\stab(\phi)|}\:,
\end{align} 
which is $5760\times(3/3+1/6)=6720$ unique SIC-POVMs in dimension 15,
for example. The stabiliser of each orbit is also given in
table~\ref{tbl:symmetries}. The numerical initial fiducial vector
$\ket{\phi}$ and its stabiliser $\stab(\phi)$ were always chosen in a
way that the stabiliser elements take the form $\cle{F}{0}$ (recall
that $\cle{F}{p}\coloneq [E_{\ce{F}{p}}]$) and, therefore, only the
matrices $F$ are quoted.\pagenote{Note that when $d$ is even, however,
  the subgroup of $\ESLtwo(\Z_{2d})$ that defines
  $\stab(\phi)\leq\PEC(d)$ can have an order that is a multiple of
  $|\stab(\phi)|$. We have chosen all stabiliser orders to double when
  treated as subgroups of $\ESLtwo(\Z_{2d})$,
  e.g.\ $|\langle\Fz\rangle|=2|\langle\cle{\Fz}{0}\rangle|$.} The
number of stabilised fiducial vectors in each orbit is also provided,
which is further divided into the number of stabilised SIC-POVMs times
the number of stabilised vectors per SIC-POVM. For example,
$\orb(\phi_{15a})$ contains 24 SIC-POVMs stabilised by
$\cle{\Fz}{0}=\matbv{0}{14}{1}{14}{0}{0}$, each containing 3
stabilised vectors; $\orb(\phi_{15d})$ contains 12 SIC-POVMs
stabilised by $\matbv{4}{11}{4}{0}{0}{0}$, each containing a single
stabilised vector.

In agreement with Appleby's analysis~\cite{Appleby05} of the previous solutions~\cite{Renes04}, in dimensions $d>3$ each stabiliser of the new solutions is a cyclic group of order a multiple of 3, and the vast majority, up to group conjugacy, have the symmetry described by Zauner's order-3 unitary $\cle{\Fz}{0}$, where (eq.~(\ref{eq:Fz}))
\begin{align}
\Fz &\coloneq \begin{pmatrix} 0 & d-1 \\ d+1 & d-1 \end{pmatrix}.
\end{align}
Exceptions occur in dimensions $d=9k+3=12(b)$, $21(e)$, $30(d)$, $39(g,h,i,j)$, $48(e,g)$, $57,\dots$, in which case solutions stabilised by the order-3 unitary $\cle{\Fsa}{0}$ exist, as indicated in parentheses, where
\begin{align}
\Fsa &\coloneq \begin{pmatrix} 1 & d+3 \\ d+3k & d-2 \end{pmatrix}.
\end{align}
The solution $\phi_{12b}$ was known previously~\cite{Grassl05}, but
the remaining are new. Following Appleby~\cite{Appleby05}, call an
order-3 (unitary) element $\cle{F}{p}\in\PEC(d)$ {\em canonical\/} if
$\tr(F)=-1\md{d}$ and $F\neq I$. Note that Zauner's unitary
$\cle{\Fz}{0}$ is canonical. In fact, it can be shown that all 
other canonical elements are conjugate to it when $9\nmid(d-3)$; 
otherwise, there are exactly two conjugacy classes of canonical 
elements in $\PEC(d)$, containing $\cle{\Fz}{0}$ and $\cle{\Fsa}{0}$, 
respectively.\pagenote{See Flammia~\cite[Conjecture~4]{Flammia06} and 
  note that $\Fz$ and $J\Fz J$ belong to different conjugacy classes 
  in $\SLtwo(\Z_{\db})$ when $3\mid d$.} Order-3 canonical unitaries 
are therefore special, in that, in the dimensions tested, every such
unitary stabilises a fiducial vector.

We can deduce two more general symmetries present in the new solutions: in dimensions $d=k^2-1=8(b)$, $15(d)$, $24(c)$, $35(i,j)$, $48(f)$, $63,\dots$, the order-2 permutation $\cle{\Fsb}{0}$, where 
\begin{align}\label{eq:symmetryb}
\Fsb &\coloneq \begin{pmatrix} -k & d \\ d & d-k \end{pmatrix},
\end{align}
stabilises the solutions indicated; in dimensions $d=(3k\pm 1)^2+3=4(a)$, $7(b)$, $19(d,e)$, $28(c)$, $52,\dots$, the order-2 anti-unitary $\cle{\Fsc}{0}$, where 
\begin{align}
\Fsc &\coloneq \begin{pmatrix} \kappa & d-2\kappa \\ d+2\kappa & d-\kappa \end{pmatrix},\quad \kappa=3k^2\pm k+1\:,
\end{align}
stabilises the solutions indicated. As with Zauner's symmetry, the reason for these extra symmetries is unknown, but their existence assists the discovery of new analytical solutions.

\afterpage{\clearpage\begin{longtable}[c]{|c|c|c|c|c|c|c|c|}
\caption{\label{tbl:symmetries}Weyl-Heisenberg covariant SIC-POVMs}\\\hline
\mr{3}{$d$}&\multirow{3}{1.35cm}{\# SIC-POVMs}&\mr{3}{$\dfrac{|\PEC(d)|}{d^2}$}&\multicolumn{5}{|c|}{$\PEC(d)$ orbits} \\\cline{4-8}
           &                                  &                                & \mr{2}{\#} & \multicolumn{2}{|c|}{stabiliser} & \multirow{2}{2.2cm}{\# stab.\ vecs.\ \hspace*{0.2cm}(per orbit)} & \mr{2}{labels} \\\cline{5-6}
           &                                  &                                &            & $|\stab|$ & $\stab$ & & \\\hline\hline
\endfirsthead
\caption{(continued)}\\\hline
\mr{3}{$d$}&\multirow{3}{1.35cm}{\# SIC-POVMs}&\mr{3}{$\dfrac{|\PEC(d)|}{d^2}$}&\multicolumn{5}{|c|}{$\PEC(d)$ orbits} \\\cline{4-8}
           &                                  &                                & \mr{2}{\#} & \multicolumn{2}{|c|}{stabiliser} & \multirow{2}{2.2cm}{\# stab.\ vecs.\ \hspace*{0.2cm}(per orbit)} & \mr{2}{labels} \\\cline{5-6}
           &                                  &                                &            & $|\stab|$ & $\stab$ & & \\\hline\hline
\endhead
\endfoot
\hline
\endlastfoot
2          & 2                 & 12                               & 1      & 6      & $\langle \mat{0}{-1}{-1}{0},\Fz \rangle$ & 2 & $a$          \\\hline
\mr{3}{3}  & \mr{3}{$\infty$}  & \mr{3}{48}                       &$\infty$& 6      & $\langle \mat{0}{-1}{-1}{0},\Fz \rangle$ & $\mspace{30mu} 2\times 3$ & $a$          \\           
           &                   &                                  & 1      & 12     & $\langle \mat{0}{-1}{-1}{0},-\Fz \rangle$ & 1 & $b$          \\           
           &                   &                                  & 1      & 48     & $\ESLtwo(\Z_3)$ & 1 & $c$          \\\hline
4          & 16                & 96                               & 1      & 6      & $\langle \Fsc\Fz \rangle = \langle \Fsc \rangle\langle \Fz \rangle$   & 2 & $a$          \\\hline
5          & 80                & 240                              & 1      & 3      & $\langle \Fz \rangle$   & 8 & $a$          \\\hline
6          & 96                & 288                              & 1      & 3      & $\langle \Fz \rangle$   & $\mspace{30mu} 12\times 3$ & $a$          \\\hline
\mr{2}{7}  & \mr{2}{336}       & \mr{2}{672}                      & 1      & 3      & $\langle \Fz \rangle$   & 8 & $a$          \\
           &                   &                                  & 1      & 6      & $\langle \Fsc\Fz \rangle = \langle \Fsc \rangle\langle \Fz \rangle$  & 2 & $b$          \\\hline
\mr{2}{8}  & \mr{2}{320}       & \mr{2}{768}                      & 1      & 3      & $\langle \Fz \rangle$   & 16 & $a$          \\
           &                   &                                  & 1      & 12     & $\langle \mat{6}{11}{5}{1} \rangle = \langle \Fz \rangle  \langle \mat{3}{6}{10}{9} \rangle\ni\Fsb$ & 2 & $b$  \\\hline
9          & 864               & 1296                             & 2      & 3      & $\langle \Fz \rangle$    & $\mspace{30mu} 12\times 3$ & $a,b$        \\\hline
10         & 480               & 1440                             & 1      & 3      & $\langle \Fz \rangle$    & 24 & $a$          \\\hline
11         & 2640              & 2640                             & 3      & 3      & $\langle \Fz \rangle$    & 16 & $a$--$c$     \\\hline
\mr{2}{12} & \mr{2}{1152}      & \mr{2}{2304}                     & 1      & 3      & $\langle \Fz \rangle$    & $\mspace{30mu} 24\times 3$ & $a$          \\
           &                   &                                  & 1      & 6      & $\langle \mat{0}{17}{17}{15} \rangle\ni\Fsa$ & $\mspace{30mu} 4\times 3$ & $b$          \\\hline
13         & 2912              & 4368                             & 2      & 3      & $\langle \Fz \rangle$    & 16 & $a,b$        \\\hline
14         & 2688              & 4032                             & 2      & 3      & $\langle \Fz \rangle$    & 24 & $a,b$        \\\hline
\mr{2}{15} & \mr{2}{6720}      & \mr{2}{5760}                     & 3      & 3      & $\langle \Fz \rangle$    & $\mspace{30mu} 24\times 3$ & $a$--$c$     \\
           &                   &                                  & 1      & 6      & $\langle \Fsb\Fz \rangle = \langle\Fsb\rangle\langle\Fz\rangle$ & 12 & $d$        \\\hline
16         & 4096              & 6144                             & 2      & 3      & $\langle \Fz \rangle$    & 32 & $a,b$        \\\hline
17         & 9792              & 9792                             & 3      & 3      & $\langle \Fz \rangle$    & 24 & $a$--$c$     \\\hline
18         & 5184              & 7776                             & 2      & 3      & $\langle \Fz \rangle$    & $\mspace{30mu} 36\times 3$ & $a,b$        \\\hline
\mr{3}{19} & \mr{3}{16720}     & \mr{3}{13680}                    & 3      & 3      & $\langle \Fz \rangle$    & 24 & $a$--$c$      \\
           &                   &                                  & 1      & 6      & $\langle \Fsc\Fz \rangle = \langle \Fsc \rangle  \langle \Fz \rangle$  & 6 & $d$          \\
           &                   &                                  & 1      & 18     & $\langle \mat{3}{12}{7}{15} \rangle = \langle \Fsc \rangle  \langle \mat{7}{14}{5}{2} \rangle \ni \Fz$ & 2 & $e$          \\\hline
20         & 7680              & 11520                            & 2      & 3      & $\langle \Fz \rangle$    & 48 & $a,b$        \\\hline
\mr{2}{21} & \mr{2}{26880}     & \mr{2}{16128}                    & 4      & 3      & $\langle \Fz \rangle$    & $\mspace{30mu} 24\times 3$ & $a$--$d$     \\
           &                   &                                  & 1      & 3      & $\langle \Fsa \rangle$ & $\mspace{30mu} 192\times 9$ & $e$          \\\hline
22         & 5280              & 15840                            & 1      & 3      & $\langle \Fz \rangle$    & 48 & $a$          \\\hline
23         & 48576             & 24288                            & 6      & 3      & $\langle \Fz \rangle$    & 32 & $a$--$f$     \\\hline
\mr{2}{24} & \mr{2}{15360}     & \mr{2}{18432}                    & 2      & 3      & $\langle \Fz \rangle$    & $\mspace{30mu} 48\times 3$ & $a,b$        \\
           &                   &                                  & 1      & 6      & $\langle \Fsb\Fz \rangle = \langle \Fsb \rangle  \langle \Fz \rangle$ & $\mspace{30mu} 24\times 3$ & $c$          \\\hline
25         & 20000             & 30000                            & 2      & 3      & $\langle \Fz \rangle$    & 40 & $a,b$        \\\hline
26         & 34944             & 26208                            & 4      & 3      & $\langle \Fz \rangle$    & 48 & $a$--$d$     \\\hline
27         & 69984             & 34992                            & 6      & 3      & $\langle \Fz \rangle$    & $\mspace{30mu} 36\times 3$ & $a$--$f$     \\\hline
\mr{2}{28} & \mr{2}{26880}     & \mr{2}{32256}                    & 2      & 3      & $\langle \Fz \rangle$    & 48 & $a,b$        \\
           &                   &                                  & 1      & 6      & $\langle \Fsc\Fz \rangle = \langle \Fsc \rangle\langle \Fz \rangle$ & 12 & $c$          \\\hline
29         & 64960             & 48720                            & 4      & 3      & $\langle \Fz \rangle$    & 40 & $a$--$d$     \\\hline\pagebreak
\mr{2}{30} & \mr{2}{46080}     & \mr{2}{34560}                    & 3      & 3      & $\langle \Fz \rangle$    & $\mspace{30mu} 72\times 3$ & $a$--$c$     \\ 
           &                   &                                  & 1      & 3      & $\langle \Fsa \rangle$ & $\mspace{30mu} 576\times 9$ & $d$          \\\hline
31         & 138880            & 59520                            & 7      & 3      & $\langle \Fz \rangle$    & 40 & $a$--$g$     \\\hline
32         & 32768             & 49152                            & 2      & 3      & $\langle \Fz \rangle$    & 64 & $a,b$        \\\hline
33         & 84480             & 63360                            & 4      & 3      & $\langle \Fz \rangle$    & $\mspace{30mu} 48\times 3$ & $a$--$d$     \\\hline
34         & 39168             & 58752                            & 2      & 3      & $\langle \Fz \rangle$    & 72 & $a,b$        \\\hline
\mr{3}{35} & \mr{3}{235200}    & \mr{3}{80640}                    & 8      & 3      & $\langle \Fz \rangle$    & 48 & $a$--$h$     \\ 
           &                   &                                  & 1      & 6      & $\langle \Fsb\Fz \rangle = \langle \Fsb \rangle  \langle \Fz \rangle$   & 24 & $i$          \\
           &                   &                                  & 1      & 12     & $\langle \mat{15}{3}{32}{18} \rangle = \langle \Fz \rangle  \langle \mat{3}{15}{20}{18} \rangle \ni \Fsb$ & 6 & $j$          \\\hline
36         & 82944             & 62208                            & 4      & 3      & $\langle \Fz \rangle$    & $\mspace{30mu} 72\times 3$ & $a$--$d$     \\\hline
37         & 134976            & 101232                           & 4      & 3      & $\langle \Fz \rangle$    & 48 & $a$--$d$     \\\hline
38         & 109440            & 82080                            & 4      & 3      & $\langle \Fz \rangle$    & 72 & $a$--$d$     \\\hline
\mr{3}{39} & \mr{3}{314496}    & \mr{3}{104832}                   & 6      & 3      & $\langle \Fz \rangle$    & $\mspace{30mu} 48\times 3$ & $a$--$f$     \\
           &                   &                                  & 2      & 3      & $\langle \Fsa \rangle$   & $\mspace{30mu} 384\times 9$ & $g,h$        \\
           &                   &                                  & 2      & 6      & $\langle \mat{0}{7}{28}{6} \rangle\ni\Fsa$ & $\mspace{30mu} 8\times 3$ & $i,j$        \\\hline
40         & 61440             & 92160                            & 2      & 3      & $\langle \Fz \rangle$    & 96 & $a,b$        \\\hline
41         & 367360            & 137760                           & 8      & 3      & $\langle \Fz \rangle$    & 56 & $a$--$h$     \\\hline
42         & 129024            & 96768                            & 4      & 3      & $\langle \Fz \rangle$    & $\mspace{30mu} 72\times 3$ & $a$--$d$     \\\hline
43         & 317856            & 158928                           & 6      & 3      & $\langle \Fz \rangle$    & 56 & $a$--$f$     \\\hline
44         & 253440            & 126720                           & 6      & 3      & $\langle \Fz \rangle$    & 96 & $a$--$f$     \\\hline
45         & 207360            & 155520                           & 4      & 3      & $\langle \Fz \rangle$    & $\mspace{30mu} 72\times 3$ & $a$--$d$     \\\hline
46         & 145728            & 145728                           & 3      & 3      & $\langle \Fz \rangle$    & 96 & $a$--$c$     \\\hline
47         & 553472            & 207552                           & 8      & 3      & $\langle \Fz \rangle$    & 64 & $a$--$h$   \\\hline
\mr{4}{48} & \mr{4}{276480}    & \mr{4}{147456}                   & 4      & 3      & $\langle \Fz \rangle$    & $\mspace{30mu} 96\times 3$ & $a$--$d$   \\
           &                   &                                  & 1      & 3      & $\langle \Fsa \rangle$   & $\mspace{30mu} 768\times 9$ & $e$   \\
           &                   &                                  & 1      & 6      & $\langle \Fsb\Fz \rangle = \langle \Fsb \rangle \langle \Fz \rangle$ & 48 & $f$   \\
           &                   &                                  & 1      & 24     & $\langle \mat{4}{37}{25}{63} \rangle\ni\Fsa,\Fsb$ & 8 & $g$   \\\hline
49         & 537824            & 230496                           & 7      & 3      & $\langle \Fz \rangle$  & 56  & $a$--$g$   \\\hline
50         & 120000            & 180000                           & 2      & 3      & $\langle \Fz \rangle$  & 120 & $a,b$      \\\hline
51--65,67  & \mr{2}{$\geq 1$}  &   \mr{2}{--}                     & \mr{2}{$\geq 1$} & \mr{2}{$\geq 3$} &  $? \ni \Fz$ & \mr{2}{--} & \mr{2}{$a$}   \\
66         &                   &                                  &  &  &  $? \ni \Fsa$      &        &             
\end{longtable}}

In table~\ref{tbl:Zsymmetries} we list the total number of fiducial vectors stabilised by Zauner's unitary matrix $\Sz\in\cle{\Fz}{0}$ in terms of fiducial-vector orbits. The number of fiducial vectors contained in each $\Sz$-eigenspace is also given ($\mathcal{Z}_k$ is as defined above eq.~(\ref{eq:Zeignenspaces})), which is further divided into the number of $\Sz$-stabilised SIC-POVMs times the number of $\Sz$-stabilised vectors per SIC-POVM. The latter is 3 when $3\mid d$, and 1 otherwise. The vectors tend to populate the largest eigenspace(s) only. This is $\mathcal{Z}_0$, except in dimensions $d=3k-1$, where $\mathcal{Z}_0$ and $\mathcal{Z}_1$ have the same largest dimension. The $\Sz$-stabilised vectors then divide evenly between these eigenspaces. Exceptions to this rule occur in dimensions $d=9k-1=8(b)$, $17(c)$, $26(d)$, $35(h,i,j)$, $44(e,f)$, $53,\dots$, where orbits exist containing only $\Sz$-stabilised vectors from $\mathcal{Z}_2$, as indicated in parentheses. 

It is of some interest whether real fiducial vectors exist~\cite{Khatirinejad08}. A search through our orbits shows that, for $d\leq 50$, these exist only in dimensions $d=3(c)$, $7(b)$, $19(e,d)$ and $39(i,j)$, of which, those in dimension 39 are new. Since $\mat{5}{29}{29}{28}\mat{0}{7}{28}{6}{}^3\mat{5}{29}{29}{28}{}^{-1}=J\md{39}$, examples from each of the two orbits can be found by multiplying the vectors given in appendix~\ref{app:numerics} by the unitary $\matbv{5}{29}{29}{28}{0}{0}$.

Also of interest are Schmidt decompositions of the fiducial vectors
for different bipartitions $\C^d\cong\C^{d_1}\otimes\C^{d_2}$, where
$d_1< d_2$ are coprime, under the identification
$\ket{k}\rightarrow\ket{k\md{d_1}}\otimes\ket{k\md{d_2}}$. This choice
preserves group covariance of SIC-POVMs by realising the group
isomorphism $\H(d)\cong\H(d_1)\times\H(d_2)$~\cite{Gross08}. Given the
singular value decomposition of the $d_1\times d_2$ matrix
$M(\phi)\coloneq\sum_{k_1,k_2}(\bra{k_1}\otimes\bra{k_2})\ket{\phi}\ket{k_1}\bra{k_2}
=\sum_l \sqrt{\lambda_l}\ket{u_l}\bra{{v_l}^*}$ (complex conjugation
in the standard basis), the Schmidt decomposition of $\ket{\phi}$ is
defined as
\begin{align}
\ket{\phi} &= \sum_l \sqrt{\lambda_l}\ket{u_l}\otimes\ket{v_l}\:,
\end{align}
where the $\lambda_l$ are called Schmidt coefficients. When $d_1=2$
(and, therefore, $d_2$ must be odd), these coefficients can be
calculated independently of the fiducial state,
$\lambda=\big(1\pm\sqrt{3/(d+1)}\big)/2$, leading to some interest in
the approach~\cite{Gross08}. This is because, along with normalisation
$\sum_l\lambda_l=1$, the Schmidt coefficients always satisfy
\begin{align}
\sum_l{\lambda_l}^2 &= \tr_1\bigl(\tr_2\ketbra{\phi}\bigr)^2 \\
&= \frac{1}{d^2}\sum_{q,p} \tr_1\,\bigl(\tr_2\, D_q^{(1)}\otimes D_p^{(2)} \ketbra{\phi}{D_q^{(1)}}^\dag\otimes {D_p^{(2)}}^\dag\bigr)^2 \\
&= \mathop{\int}_{\PCd}\d\mu(x)\,\tr_1\bigl(\tr_2\ketbra{x}\bigr)^2 \\
&= \frac{d_1+d_2}{d+1}\:,
\end{align}
using group covariance (where the displacement $D_q^{(k)}$ acts on $\C^{d_k}$), the property of 2-designs [eq.~(\ref{eq:tdesign})], and ref.~\cite{Lubkin78} to evaluate the integral. Some other special Schmidt coefficients are apparent from our solutions (to the numerical precision given): when $d_1=3$, we find that  $\lambda=0,\big(1\pm 1/\sqrt{13}\big)/2$ in dimension $d=12(b)$, $\lambda=0,1/2,1/2$ in dimension $d=15(d)$, and $\lambda=1/7,3/7,3/7$ in dimension $d=48(f,g)$; when $d_1=5$, we find $\lambda=0,0,1/3,1/3,1/3$ in dimension $d=35(i,j)$. Nice Schmidt coefficients are seen to follow from the $\Fsb$-symmetry [eq.~(\ref{eq:symmetryb})] except for $d=24(c)$, where we deduce that $\lambda=\big(\sqrt{21}-1\big)/10$, $\big(11-\sqrt{21}\pm\sqrt{26\sqrt{21}-98}\big)/20$, when $d_1=3$.

\afterpage{\clearpage\begin{longtable}[c]{|c|c|c|c|c|c|c|c|c|}
\caption{\label{tbl:Zsymmetries}Fiducial vectors with Zauner symmetry}\\\hline
\mr{3}{$d$}&\multirow{3}{2.6cm}{\# $\Sz$-stabilised fiducial vectors}&\multicolumn{7}{|c|}{$\PEC(d)$ orbits} \\\cline{3-9}
           &                                  & \mr{2}{\#} & \mr{2}{$|\stab|$} & \multicolumn{4}{|c|}{\# $\Sz$-stabilised vectors (per orbit)} & \mr{2}{labels} \\\cline{5-8}
           &                                  &            &                      &$\mspace{60mu}$total$\mspace{60mu}$& $\mspace{40mu}\mathcal{Z}_0\mspace{40mu}$ & $\mspace{40mu}\mathcal{Z}_1\mspace{40mu}$ & $\mspace{40mu}\mathcal{Z}_2\mspace{40mu}$ & \\\hline\hline
\endfirsthead
\caption{(continued)}\\\hline
\mr{3}{$d$}&\multirow{3}{2.6cm}{\# $\Sz$-stabilised fiducial vectors}&\multicolumn{7}{|c|}{$\PEC(d)$ orbits} \\\cline{3-9}
           &                                  & \mr{2}{\#} & \mr{2}{$|\stab|$} & \multicolumn{4}{|c|}{\# $\Sz$-stabilised vectors (per orbit)} & \mr{2}{labels} \\\cline{5-8}
           &                                  &            &                      &$\mspace{60mu}$total$\mspace{60mu}$& $\mspace{40mu}\mathcal{Z}_0\mspace{40mu}$ & $\mspace{40mu}\mathcal{Z}_1\mspace{40mu}$ & $\mspace{40mu}\mathcal{Z}_2\mspace{40mu}$ & \\\hline\hline
\endhead
\endfoot
\hline
\endlastfoot
2          & 2                                & 1      & 6      & 2  & 1  & 1  & 0  & $a$          \\\hline
\mr{3}{3}  & \mr{3}{$\infty$}                 &$\infty$& 6      & $2\rlap{${}\times 3$}$  & $2\rlap{${}\times 3$}$  & 0  & 0  & $a$          \\           
           &                                  & 1      & 12     & $1\rlap{${}\times 3$}$  & $1\rlap{${}\times 3$}$  & 0  & 0  & $b$          \\           
           &                                  & 1      & 48     & $1\rlap{${}\times 3$}$  & $1\rlap{${}\times 3$}$  & 0  & 0  & $c$          \\\hline
4          & 4                                & 1      & 6      & $4$  & $4$  & $0$  & $0$  & $a$          \\\hline
5          & 8                                & 1      & 3      & 8  & 4  & 4  & 0  & $a$          \\\hline
6          & $12\rlap{${}\times 3$}$        & 1      & 3      & $12\rlap{${}\times 3$}$ & $12\rlap{${}\times 3$}$ & 0  & 0  & $a$          \\\hline
\mr{2}{7}  & \mr{2}{12}                       & 1      & 3      & 8  & 8  & 0  & 0  & $a$          \\
           &                                  & 1      & 6      & 4  & 4  & 0  & 0  & $b$          \\\hline
\mr{2}{8}  & \mr{2}{20}                       & 1      & 3      & 16 & 8  & 8  & 0  & $a$          \\
           &                                  & 1      & 12     & 4  & 0  & 0  & 4  & $b$  \\\hline
9          & $24\rlap{${}\times 3$}$        & 2      & 3      & $12\rlap{${}\times 3$}$ & $12\rlap{${}\times 3$}$ & 0  & 0  & $a,b$        \\\hline
10         & 24                               & 1      & 3      & 24 & 24 & 0  & 0  & $a$          \\\hline
11         & 48                               & 3      & 3      & 16 & 8  & 8  & 0  & $a$--$c$     \\\hline
\mr{2}{12} & \mr{2}{$24\rlap{${}\times 3$}$}& 1      & 3      & $24\rlap{${}\times 3$}$ & $24\rlap{${}\times 3$}$ & 0  & 0  & $a$          \\
           &                                  & 1      & 6      & 0  & 0  & 0  & 0  & $b$          \\\hline
13         & 32                               & 2      & 3      & 16 & 16 & 0  & 0  & $a,b$        \\\hline
14         & 48                               & 2      & 3      & 24 & 12 & 12 & 0  & $a,b$        \\\hline
\mr{2}{15} & \mr{2}{$84\rlap{${}\times 3$}$}& 3      & 3      & $24\rlap{${}\times 3$}$ & $24\rlap{${}\times 3$}$ & 0  & 0  & $a$--$c$     \\
           &                                  & 1      & 6      & $12\rlap{${}\times 3$}$ & $12\rlap{${}\times 3$}$ & 0  & 0  & $d$        \\\hline
16         & 64                               & 2      & 3      & 32 & 32 & 0  & 0  & $a,b$        \\\hline
\mr{2}{17} & \mr{2}{72}                       & 2      & 3      & 24 & 12 & 12 & 0  & $a,b$     \\
           &                                  & 1      & 3      & 24 & 0  & 0  & 24 & $c$     \\\hline
18         & $72\rlap{${}\times 3$}$        & 2      & 3      & $36\rlap{${}\times 3$}$ & $36\rlap{${}\times 3$}$ & 0  & 0   & $a,b$        \\\hline
\mr{3}{19} & \mr{3}{88}                       & 3      & 3      & 24 & 24 & 0  & 0  & $a$--$c$      \\
           &                                  & 1      & 6      & 12 & 12 & 0  & 0 & $d$          \\
           &                                  & 1      & 18     & 4  & 4  & 0  & 0 & $e$          \\\hline
20         & 96                               & 2      & 3      & 48 & 24 & 24 & 0  & $a,b$        \\\hline
\mr{2}{21} & \mr{2}{$96\rlap{${}\times 3$}$}& 4      & 3      & $24\rlap{${}\times 3$}$ & $24\rlap{${}\times 3$}$ & 0  & 0  & $a$--$d$     \\
           &                                  & 1      & 3      & 0  & 0  & 0  & 0    & $e$          \\\hline
22         & 48                               & 1      & 3      & 48 & 48 & 0  & 0  & $a$          \\\hline
23         & 192                              & 6      & 3      & 32 & 16 & 16 & 0  & $a$--$f$     \\\hline
\mr{2}{24} &\mr{2}{$120\rlap{${}\times 3$}$}& 2      & 3      & $48\rlap{${}\times 3$}$ & $48\rlap{${}\times 3$}$ & 0  & 0   & $a,b$        \\
           &                                  & 1      & 6      & $24\rlap{${}\times 3$}$ & $24\rlap{${}\times 3$}$ & 0  & 0  & $c$          \\\hline
25         & 80                               & 2      & 3      & 40 & 40 & 0  & 0  & $a,b$        \\\hline
\mr{2}{26} & \mr{2}{192}                      & 3      & 3      & 48 & 24 & 24 & 0  & $a$--$c$     \\
           &                                  & 1      & 3      & 48 & 0  & 0  & 48 & $d$     \\\hline
27         & $216\rlap{${}\times 3$}$       & 6      & 3      & $36\rlap{${}\times 3$}$ & $36\rlap{${}\times 3$}$ & 0  & 0   & $a$--$f$     \\\hline
\mr{2}{28} & \mr{2}{120}                      & 2      & 3      & 48 & 48 & 0  & 0  & $a,b$        \\
           &                                  & 1      & 6      & 24 & 24 & 0  & 0  & $c$          \\\hline
29         & 160                              & 4      & 3      & 40 & 20 & 20 & 0  & $a$--$d$     \\\hline
\mr{2}{30} &\mr{2}{$216\rlap{${}\times 3$}$}& 3      & 3      & $72\rlap{${}\times 3$}$ & $72\rlap{${}\times 3$}$ & 0  & 0   & $a$--$c$     \\ 
           &                                  & 1      & 3      & 0  & 0  & 0  & 0    & $d$          \\\hline
31         & 280                              & 7      & 3      & 40 & 40 & 0  & 0  & $a$--$g$     \\\hline
32         & 128                              & 2      & 3      & 64 & 32 & 32 & 0  & $a,b$        \\\hline
33         & $192\rlap{${}\times 3$}$       & 4      & 3      & $48\rlap{${}\times 3$}$ & $48\rlap{${}\times 3$}$ & 0  & 0   & $a$--$d$     \\\hline
34         & 144                              & 2      & 3      & 72 & 72 & 0  & 0  & $a,b$        \\\hline
\mr{4}{35} & \mr{4}{420}                      & 7      & 3      & 48 & 24 & 24 & 0  & $a$--$g$     \\ 
           &                                  & 1      & 3      & 48 & 0  & 0  & 48 & $h$     \\ 
           &                                  & 1      & 6      & 24 & 0  & 0  & 24 & $i$          \\
           &                                  & 1      & 12     & 12 & 0  & 0  & 12 & $j$          \\\hline
36         & $288\rlap{${}\times 3$}$       & 4      & 3      & $72\rlap{${}\times 3$}$ & $72\rlap{${}\times 3$}$ & 0  & 0   & $a$--$d$     \\\hline
37         & 192                              & 4      & 3      & 48 & 48 & 0  & 0  & $a$--$d$     \\\hline
38         & 288                              & 4      & 3      & 72 & 36 & 36 & 0  & $a$--$d$     \\\hline
\mr{3}{39} &\mr{3}{$288\rlap{${}\times 3$}$}& 6      & 3      & $48\rlap{${}\times 3$}$ & $48\rlap{${}\times 3$}$ & 0  & 0   & $a$--$f$     \\
           &                                  & 2      & 3      & 0  & 0  & 0  & 0    & $g,h$        \\
           &                                  & 2      & 6      & 0  & 0  & 0  & 0  & $i,j$        \\\hline
40         & 192                              & 2      & 3      & 96 & 96 & 0  & 0  & $a,b$        \\\hline
41         & 448                              & 8      & 3      & 56 & 28 & 28 & 0  & $a$--$h$     \\\hline
42         & $288\rlap{${}\times 3$}$       & 4      & 3      & $72\rlap{${}\times 3$}$ & $72\rlap{${}\times 3$}$ & 0  & 0   & $a$--$d$     \\\hline
43         & 336                              & 6      & 3      & 56 & 56 & 0  & 0  & $a$--$f$     \\\hline
\mr{2}{44} & \mr{2}{576}                      & 4      & 3      & 96 & 48 & 48 & 0  & $a$--$d$     \\
           &                                  & 2      & 3      & 96 & 0  & 0  & 96 & $e,f$     \\\hline
45         & $288\rlap{${}\times 3$}$       & 4      & 3      & $72\rlap{${}\times 3$}$ & $72\rlap{${}\times 3$}$ & 0  & 0   & $a$--$d$     \\\hline
46         & 288                              & 3      & 3      & 96 & 96 & 0  & 0  & $a$--$c$   \\\hline
47         & 512                              & 8      & 3      & 64 & 32 & 32 & 0  & $a$--$h$   \\\hline
\mr{4}{48} &\mr{4}{$432\rlap{${}\times 3$}$}& 4      & 3      & $96\rlap{${}\times 3$}$ & $96\rlap{${}\times 3$}$ & 0  & 0  & $a$--$d$   \\
           &                                  & 1      & 3      & 0  & 0  & 0  & 0  & $e$   \\
           &                                  & 1      & 6      & $48\rlap{${}\times 3$}$ & $48\rlap{${}\times 3$}$ & 0  & 0  & $f$   \\
           &                                  & 1      & 24     & 0  & 0  & 0  & 0  & $g$   \\\hline
49         & 392                              & 7      & 3      & 56 & 56 & 0  & 0      & $a$--$g$      \\\hline
50         & 240                              & 2      & 3      & 120& 60 & 60 & 0      & $a,b$      
\end{longtable}}

\section{Symbolical computer solutions}
\label{sec:symb}
Algebraic solutions for Weyl-Heisenberg SIC-POVMs have been reported
by Zauner~\cite{Zauner99} for dimensions $d=2,3,4,5$, together with a
solution for $d=8$ due to Hoggar \cite{Hoggar98} which is
covariant with respect to the three-qubit Pauli group.
Appleby~\cite{Appleby05} added Weyl-Heisenberg solutions for $d=7$ and
$d=19$.  The second author has reported on algebraic solutions for
dimension $d=6,\ldots,13$ and $d=15$ in a series of conference talks
\cite{Grassl04,Grassl05,Grassl06,Grassl08a,Grassl08b}.  Here we close
the gap at $d=14$ and add new solutions for $d=24,35,48$.

The general approach used for finding algebraic solutions is to solve
the system of polynomial equations for a fiducial vector that lies in
an eigenspace of a prescribed symmetry, for example, the Zauner matrix
(see e.g.\ \cite{Grassl08b} for more details).  The smaller the
dimension of the eigenspace, the fewer the variables that are needed, and the
higher the chances are of being able to compute a solution.  The new
solutions for $d=24,35,48$ have been obtained with the help
of the comparatively large symmetry groups identified in these dimensions,
upon inspection of the numerical solutions.

In order to be able to express the Hermitian inner product and the
squared modulus via polynomials, we have to replace each complex
variable $c_j=\braket{j}{\phi}$ by two real variables, i.e., $c_j=a_j+i b_j$ where
$i^2=-1$.  In general, the resulting system of polynomial equations
will have both real and complex solutions for the variables $a_j$ and
$b_j$, but we are only interested in those solutions for which all
components are real.

Note that in an algebraic extension of the rationals $\Q$, there is no
\textit{a priori} notion of an element being \emph{positive} or
\emph{real}. An element $\alpha$, for example, which is defined via
$\alpha^2-2=0$ can either be mapped to $\sqrt{2}$ or $-\sqrt{2}$.
Furthermore, the roots of $f(z)=z^4-2z^2-1$ are
$\beta\coloneq\pm\sqrt{1\pm\alpha}$. Depending on the choice of the sign of
$\alpha=\pm\sqrt{2}$, $\beta$ is mapped to a real or a complex number.
Now assume that we are given a number field $\K=\Q(\theta)$ where
$\theta$ is a root of an irreducible polynomial $f_{\K}(z)\in\Q[z]$ of
degree $2m$ with rational coefficients.  Without loss of generality,
we additionally assume that $\K$ contains a square root of $-1$.  We
can then identify a subfield $\LL\le\K$ such that $\K=\LL(\tau)$ with
$\tau^2=-1$ and $\LL=\Q(\theta_1)$ with $f_{\LL}(\theta_1)=0$ for an
irreducible polynomial $f_{\LL}\in\Q[z]$ of degree $m$ that has at
least one real root.  A field $\LL$ with this property is said to have
a real embedding.  Fixing such a representation of $\K$, we can
identify the elements of $\LL$ with the real numbers, and complex
conjugation corresponds to an automorphism $\gamma_c$ of $\K$ that
maps $\tau$ to $-\tau$ and fixes all elements of $\LL$.  As we will
illustrate below, a field automorphism of $\K$ that commutes with the
``complex conjugation'' $\gamma_c$ and stabilises the Weyl-Heisenberg
group as a set, maps a Weyl-Heisenberg SIC-POVM to a possibly
different Weyl-Heisenberg SIC-POVM.

Using the computer algebra system {\sc Magma} \cite{Magma}, we have
computed at least one fiducial vector for dimensions
$d=4,\dots,15,19,24,35,48$.  The symbolical fiducial vectors can be
found in appendix~\ref{app:symbolics}.  While the labeling of the
orbits is the same as that introduced for the numerical fiducial
vectors, the vectors themselves are not necessarily the very same as
those given in appendix~\ref{app:numerics}. Their lengths are also
unnormalised.  Note that in addition to basic arithmetic operations we
need only to compute square roots, and third or fifth roots.  This is
related to the fact that in all cases for which we have been able to
compute a solution, the solution can be found in a number field with
solvable Galois group, and hence the field extension can be expressed
in terms of radicals.\pagenote{When expressing the elements of the
  original number field in terms of radicals, we generally have to
  extend the field to one of larger degree.  The information given in
  table~\ref{tbl:fields} is with respect to the original field.}

As an intermediate step, we compute a so-called Gr\"obner basis
$\mathcal{G}$ of the ideal generated by the system of polynomial
equations. Unfortunately, the polynomials in $\mathcal{G}$ tend to
have large coefficients, and in some cases even a couple of hundred
digits, e.g.\ for $d=11$.  Likewise, the defining polynomials of the
number field containing a solution quite often have large degrees and
huge coefficients.  In most of the cases, we succeeded in computing a
more compact representation of the solution based on the non-unique
decomposition of the field. In Table~\ref{tbl:fields} we give the
degree of the number field over which a fiducial vector has been found
and which contains the square root $\tau$ of $-1$ as canonical
generator for the extension $\C=\R(\tau)$.  Note that this field does
not necessarily contain a primitive $d$-th root of unity.  One such
example is $d=19$, where the extension degree is $8$, but a $19$-th
root of unity is defined by an irreducible polynomial of degree $18$.
Hence the overall extension degree is at least $\mathop{\rm
  lcm}(8,18)=72$.  In some other cases, the field is not closed under
all possible automorphisms induced by mapping one root of the defining
polynomial to another.  In those cases, the field has to be extended
to obtain a normal field, and the Galois group is larger than the
degree of the original field.  In any case, the Galois group is
independent of the chosen representation of the field.

\begin{table}[hbt]
\begin{center}
\begin{tabular}{|c|c|c|c|}
\hline
$d$ & extension field degree& Galois group & labels\\
\hline\hline
$4$ & $16=2^4$& $\text{\texttt{SmallGroup(16,11)}}=C_2\times D_8$ &$a$\\
\hline
$5$ & $32=2^5$& $\text{\texttt{SmallGroup(32,38)}}=(C_8\times C_2)\rtimes C_2$& $a$\\
\hline
$6$ & $48=2^4\times 3$& $\text{\texttt{SmallGroup(48,43)}}=C_2\times ((C_6\times C_2)\rtimes C_2)$& $a$\\
\hline
$7$ & $8=2^3$& $\text{\texttt{SmallGroup(16,11)}}=C_2\times D_8$& $a,b$\\
\hline
\mr{2}{$8$} & $128=2^7$& $\text{\texttt{SmallGroup(128,1032)}}=C_4\times ((C_4\times C_2^2)\rtimes C_2)$& $a$\\
    & $32=2^5$& $\text{\texttt{SmallGroup(32,25)}}=C_4\times D_8$& $b$\\
\hline
$9$ & $144=2^4\times 3^2$ & $\text{\texttt{SmallGroup(144,161)}}=C_3\times ((C_{12}\times C_2) \rtimes C_2)$ & $a,b$\\
\hline
$10$ &$192=2^6\times 3$ &$\text{\texttt{SmallGroup(192,1297)}}=C_2\times ((C_{24}\times C_2) \rtimes C_2)$& $a$\\
\hline
$11$ &$320=2^6\times 5$ &$\text{\texttt{SmallGroup(320,1574)}}=C_{10}\times((C_8\times C_2)\rtimes C_2)$ & $a,b,c$\\
\hline
\mr{2}{$12$} & $192=2^6\times 3$& $\text{\texttt{SmallGroup(192,1399)}}=C_2\times((C_6\times C_2^3)\rtimes C_2)$& $a$\\
             & $32=2^5$ & $\text{\texttt{SmallGroup(64,202)}}=C_2\times (C_2^4\rtimes C_2)$& $b$\\
\hline
$13$ & $32=2^5$& $\text{\texttt{SmallGroup(128,1685)}}=C_2\times ((C_8\times C_4)\rtimes C_2)$& $a,b$\\
\hline
\mr{2}{$14$} &\mr{2}{$96=2^5\times 3$}&$\text{\texttt{SmallGroup(12,5)}}\times\text{\texttt{SmallGroup(768,202015)}}$ & \mr{2}{$a,b$}\\
  &&${}=C_2\times C_6\times ((C_6\times ((C_2\times ((C_4\times C_2)\rtimes C_2))\rtimes C_2))\rtimes C_2)$ & \\
\hline
$15$ &$96=2^5\times 3$ & $\text{\texttt{SmallGroup(96,108)}}=(C_{24}\times C_2)\rtimes C_2$& $d$\\
\hline\hline
$19$ & $8=2^3$& $\text{\texttt{SmallGroup(16,11)}}=C_2\times D_8$& $e$\\
\hline
$24$ & $384=2^7\times 3$&$\text{\texttt{SmallGroup(768,327907)}}=C_2\times ((C_{12} \times C_4\times C_2\times C_2)\rtimes C_2)$  & $c$\\
\hline
$35$ & $576=2^6\times 3^2$&$\text{\texttt{SmallGroup(576,7767)}}=C_2\times C_{12}\times((C_6\times C_2)\rtimes C_2)$& $j$\\
\hline
\mr{2}{$48$} &\mr{2}{$512=2^9$} & $\text{\texttt{SmallGroup(16,5)}}\times\text{\texttt{SmallGroup(64,146)}}$&\mr{2}{$g$}\\
             &                  & ${}=C_2\times C_8\times ((C_8\times C_2\times C_2)\rtimes C_2)$&\\
\hline
\end{tabular}
\end{center}
\caption{Degree of the number field in which a fiducial vector for
  dimension $d$ has been found.  The Galois group of a normal field
  extension is given as an additional invariant of the
  field.\label{tbl:fields}}
\end{table}

The representation of the fiducial vector, and thereby the SIC-POVM,
by algebraic numbers allowed us to identify a relation between the two
orbits under the extended Clifford group for $d=9$.  Fiducial vectors
of both orbits labeled $9a$ and $9b$ can be found in an eigenspace of
the Zauner matrix, and they are defined over the same number field
\begin{align}
\K_9 &= \Q\Bigl(\sqrt{3},\sqrt{5},\sigma_1,\sigma_2,\sigma_3=\sqrt{\sqrt{15}/2+\sqrt{3}},\tau=\sqrt{-1}\Bigr)\:,
\end{align}
where $\sigma_1$ and $\sigma_2$ are roots of the polynomials
$f_1(z)=z^3-3z-1$ and $f_2(z)=z^3-18z+12$, respectively.  It turns out
that the following automorphism of $\K_9$ connects the two orbits:
\begin{align}
\gamma &\colon \sqrt{3}\mapsto -\sqrt{3}\:,\;\sqrt{5}\mapsto-\sqrt{5}\:,\;\sigma_3\mapsto(\sqrt{5}-2)\sigma_3\:,
\end{align}
and fixes $\sigma_1$, $\sigma_2$ and $\tau$. Note that the change of
signs of the square roots of $3$ and $5$ implies the mapping for
$\sigma_3$, since the defining polynomial of $\sigma_3$ then changes
as well.  The set of triple products
$C_{jkl}=\braket{x_j}{x_k}\braket{x_k}{x_l}\braket{x_l}{x_j}$ computed
for all triples of states in a single Weyl-Heisenberg orbit, however,
is not invariant under $\gamma$.  This implies that the two orbits are
not related by a unitary or anti-unitary transformation (see
ref.~\cite{Appleby08}). Similarly, the two orbits $13a$ and $13b$ are
related by a field automorphism of order $8$, implied by
simultaneously changing the signs of $\sqrt{5}$ and $\sqrt{7}$.  For
dimension $11$, the orbits $11a$ and $11b$ are related by a field
automorphism of order $8$, implied by simultaneously changing the
signs of $\sqrt{2}$ and $\sqrt{3}$, while the orbit $11c$ is unrelated
to them.  The orbits $14a$ and $14b$ are related by simultaneously
changing the signs of $\sqrt{3}$ and $\sqrt{5}$.

It is not yet clear to us when such a relation exists.
Unlike complex conjugation, which is intrinsic to the problem and
captured by the extended Clifford group, we cannot predict which other
field automorphisms relate the different orbits since we know the
relevant number field only after having found a solution.

\section{Conclusion}
\label{sec:end}
Ten years have now passed since it was first conjectured, by Gerhard
Zauner~\cite{Zauner99}, that maximal sets of $d^2$ equiangular lines
exist in all finite complex dimensions $d$. Despite considerable
attention now attracted to this question from the quantum physics
community, where sets of $d^2$ equiangular lines define
SIC-POVMs~\cite{Renes04}, Zauner's conjecture remains open to this
day.

The primary purpose of this article is to report the results of the
authors' new computer study on SIC-POVMs: numerical solutions for
fiducial vectors that generate Weyl-Heisenberg covariant SIC-POVMs are
provided in appendix~\ref{app:numerics} for all $d\leq 67$. These
confirm Zauner's conjecture in all such dimensions, except $d=66$,
which remains inconclusive. Moreover, a putatively complete list of
solutions is given for $d\leq 50$. It is our hope that this list will
act as an important resource for the growing community of researchers
interested in the SIC-POVM problem. A preliminary symmetry analysis
has lead to new algebraic solutions in dimensions $d=24$, $35$ and
$48$. These are collected with other known Weyl-Heisenberg covariant
algebraic solutions in appendix~\ref{app:symbolics} (see also
ref.~\cite{Appleby05}). Additional symmetries are likely to exist in
arbitrarily large dimensions, and might help in proving the existence
of an infinite series of corresponding SIC-POVMs.

Although our confidence in its truth has grown considerably, we seem
no closer to a proof of Zauner's conjecture than Gerhard Zauner was at
the time of his doctoral dissertation. The observation by Zauner, and
subsequently by others, of the apparent deep connection between the
Heisenberg group and maximal sets of equiangular lines, appears to be
born more out of luck than discernment. Our incomplete understanding
of the automorphism group of Weyl-Heisenberg covariant SIC-POVMs is
most impoverishing: simplifying to prime dimensions, although the
outer automorphism group of $\H(d)$ is $\GL_2(\Z_d)$, only members of
the subgroup $\ESLtwo(\Z_d)$ are currently known to naturally define
automorphisms of Weyl-Heisenberg SIC-POVMs (as described in
sec.~\ref{sec:whsics}). The relation found in dimension $d=9$ between
the two different extended Clifford orbits of solutions (labeled $9a$
and $9b$) hints of a bigger picture that could paint currently
unrelated orbits in the same light (see sec.~\ref{sec:symb}). Without
a deeper analysis of the known SIC-POVM solutions, however, we can
only speculate on what this bigger picture might be.

\begin{acknowledgments}
The authors acknowledge the hospitality of the Perimeter Institute 
during the workshop Seeking SICs and thank the organisers Chris Fuchs
and Steve Flammia for providing an excellent environment to exchange
ideas. We would also like to thank Marcus Appleby, David Gross, 
Gerhard Zauner and Huangjun Zhu for fruitful discussions.

Centre for Quantum Technologies is a Research Centre of Excellence
funded by Ministry of Education and National Research Foundation of
Singapore. AJS is supported by Australian Research Council grant
CE0348250 and thanks the Centre for Quantum Technologies for their
hospitality.
\end{acknowledgments}

\section{Notes}
\renewcommand*{\notedivision}{}
\renewcommand*{\pagenotesubhead}[2]{}
\printnotes

\newpage
\appendix
\section{Numerical solutions}
\label{app:numerics}

\printfid{2a}
\printfid{3a}\printfid{3b}\printfid{3c}
\printfid{4a}
\printfid{5a}
\printfid{6a}
\printfid{7a}\printfid{7b}
\printfid{8a}\printfid{8b}
\printfid{9a}\printfid{9b}
\printfid{10a}
\printfid{11a}\printfid{11b}\printfid{11c}
\printfid{12a}\printfid{12b}
\printfid{13a}\printfid{13b}
\printfid{14a}\printfid{14b}
\printfid{15a}\printfid{15b}\printfid{15c}\printfid{15d}
\printfid{16a}\printfid{16b}
\printfid{17a}\printfid{17b}\printfid{17c}
\printfid{18a}\printfid{18b}
\printfid{19a}\printfid{19b}\printfid{19c}\printfid{19d}\printfid{19e}
\printfid{20a}\printfid{20b}
\printfid{21a}\printfid{21b}\printfid{21c}\printfid{21d}\printfid{21e}
\printfid{22a}
\printfid{23a}\printfid{23b}\printfid{23c}\printfid{23d}\printfid{23e}\printfid{23f}
\printfid{24a}\printfid{24b}\printfid{24c}
\printfid{25a}\printfid{25b}
\printfid{26a}\printfid{26b}\printfid{26c}\printfid{26d}
\printfid{27a}\printfid{27b}\printfid{27c}\printfid{27d}\printfid{27e}\printfid{27f}
\printfid{28a}\printfid{28b}\printfid{28c}
\printfid{29a}\printfid{29b}\printfid{29c}\printfid{29d}
\printfid{30a}\printfid{30b}\printfid{30c}\printfid{30d}
\printfid{31a}\printfid{31b}\printfid{31c}\printfid{31d}\printfid{31e}\printfid{31f}\printfid{31g}
\printfid{32a}\printfid{32b}
\printfid{33a}\printfid{33b}\printfid{33c}\printfid{33d}
\printfid{34a}\printfid{34b}
\printfid{35a}\printfid{35b}\printfid{35c}\printfid{35d}\printfid{35e}\printfid{35f}\printfid{35g}\printfid{35h}\printfid{35i}\printfid{35j}
\printfid{36a}\printfid{36b}\printfid{36c}\printfid{36d}
\printfid{37a}\printfid{37b}\printfid{37c}\printfid{37d}
\printfid{38a}\printfid{38b}\printfid{38c}\printfid{38d}
\printfid{39a}\printfid{39b}\printfid{39c}\printfid{39d}\printfid{39e}\printfid{39f}\printfid{39g}\printfid{39h}\printfid{39i}\printfid{39j}
\printfid{40a}\printfid{40b}
\printfid{41a}\printfid{41b}\printfid{41c}\printfid{41d}\printfid{41e}\printfid{41f}\printfid{41g}\printfid{41h}
\printfid{42a}\printfid{42b}\printfid{42c}\printfid{42d}
\printfid{43a}\printfid{43b}\printfid{43c}\printfid{43d}\printfid{43e}\printfid{43f}
\printfid{44a}\printfid{44b}\printfid{44c}\printfid{44d}\printfid{44e}\printfid{44f}
\printfid{45a}\printfid{45b}\printfid{45c}\printfid{45d}
\printfid{46a}\printfid{46b}\printfid{46c}
\printfid{47a}\printfid{47b}\printfid{47c}\printfid{47d}\printfid{47e}\printfid{47f}\printfid{47g}\printfid{47h}
\printfid{48a}\printfid{48b}\printfid{48c}\printfid{48d}\printfid{48e}\printfid{48f}\printfid{48g}
\printfid{49a}\printfid{49b}\printfid{49c}\printfid{49d}\printfid{49e}\printfid{49f}\printfid{49g}
\printfid{50a}\printfid{50b}
\printfid{51a}
\printfid{52a}
\printfid{53a}
\printfid{54a}
\printfid{55a}
\printfid{56a}
\printfid{57a}
\printfid{58a}
\printfid{59a}
\printfid{60a}
\printfid{61a}
\printfid{62a}
\printfid{63a}
\printfid{64a}
\printfid{65a}
\printfid{66a}
\printfid{67a}

\newpage

\section{Symbolical solutions}
\label{app:symbolics}
\printfidsym{4a}
\printfidsym{5a}
\printfidsym{6a}
\printfidsym{7a}\printfidsym{7b}
\printfidsym{8a}\printfidsym{8b}
\printfidsym{9a}\printfidsym{9b}
\printfidsym{10a}
\printfidsym{11a}\printfidsym{11b}\printfidsym{11c}
\printfidsym{12a}\printfidsym{12b}
\printfidsym{13a}\printfidsym{13b}
\printfidsym{14a}\printfidsym{14b}
\printfidsym{15d}
\printfidsym{19e}
\printfidsym{24c}
\printfidsym{35j}
\printfidsym{48g}

\end{document}